\documentstyle[12pt,epsfig]{article}

\title{CONTRIBUTION OF THE ELASTIC RADIATIVE TAIL 
TO THE DEEP INELASTIC MUON
SCATTERING ON HEAVY TARGETS}
\author{Krzysztof Kurek$^*$}
\begin{document}
\date{}

\maketitle
\noindent
$^*$ {\it Institute of Physics, Warsaw University Branch, Lipowa 41,
15-424 Bia\l{}ystok, Poland,\\
.                   e-mail: kurek@fuw.edu.pl}\\

\medskip\medskip\medskip
\vskip1cm
\begin{abstract}
Improvement of the radiative correction scheme for the deep inelastic
scattering on the heavy targets is discussed. Arguments that the
contribution of the radiative tail from the
elastic peak
should be calculated without the use of the Born approximation in the
case of the heavy target scattering are presented. The semianalytical
approach based on the classical solution of the old
bremsstrahlung problem is shortly described. The numerical results for
the new correction factor $\eta$ are presented for the scattering on tin
and lead.
\end{abstract}
\medskip \medskip \medskip
\section{Introduction}

\medskip
The determination of the one photon cross section from the data which is
the goal of the electroproduction experiment demands excluding of a contribution
from the other electroweak processes. These other contributions must be
calculated theoretically and subtracted from the measured cross section.
The magnitude of these radiative effects in the measured cross
section is characterized by the radiative correction factor, $\eta(x_{Bj},y)$,
 defined as follows: 
\begin{equation}
{\eta (x_{Bj},y)} = {\sigma _{1\gamma }\over \sigma_{meas}}\
\end{equation}
The set of theoretical calculations which allows to calculate this factor 
$\eta$,
 used then in the analysis of experimental data, 
is called {\it radiative corrections scheme}.
The precision of measurements in the recent experiments is very high 
and also the estimations of the radiative
corrections have to be based on more precise theoretical calculations. 
\medskip
It this paper we shall refer to the very precise deep inelastic muon scattering
experiment
at CERN: New Muon Collaboration (NMC) experiment \cite{NMC}. All formulae
presented this the paper are also true for electron scattering experiments
and therefore we shall use the name lepton instead of muon in the 
further considerations.
Two different radiative corrections schemes have been
used in the NMC data analysis: the old one, so called Mo and Tsai scheme, 
created by L.W. Mo and Y.S. Tsai \cite{MT} and
Dubna scheme, created by A.A. Akhundov, D.Yu. Bardin, 
N.M. Shumeiko et al.
\cite{Dubna}. A more detailed description and
comparison of the MT and D schemes is presented in ref. \cite{my}.

\medskip
The goal of this paper is to estimate the contribution of the multiphotons
exchange processes in the so called {\it elastic tail}.
These processes can
be important for the heavy target scattering and were not considered
in the schemes mentioned above.

\medskip
It is well known that the one of the more important contribution to the
cross section measured in the inclusive deep inelastic experiment proceeds
from {\it elastic tail} - the bremsstrahlung process from lepton
current in the elastic channel. Especially in the low $x_{Bj}$ and high
$y$ ($y = \nu/E$) region this contribution can be dominant.
In the both schemes: MT and D, the corrections from elastic tail
are taken into account in the same way
and only in the Born approximation
\footnote{In the D scheme the Born cross section for the elastic tail 
contains also the $Z^0$ boson exchange contribution which should be in fact 
taken into account in the framework of the standard model of the electroweak 
interactions. In the practical application to the analysis of the data in
the present experiments this contributions is negligible.}.
The relevent Feynman diagrams for the bremsstrahlung coresponding to the elastic radiative tail
in the Born approximations are presented in fig.~1. To find the corresponding 
formulae
the reader is reffered to the original papers \cite{MT,Dubna,my}.
This contribution is of the order of $Z \alpha^3 $ where $\alpha$ is the
electromagnetic coupling constant ($\alpha = e^2/4\pi$) and $Ze$ is the electric
charge of the target particle.
The higher order corrections to the radiative elastic tail ($Z \alpha^4 $ and
higher) have been also estimated in the one photon exchange approximation
and shown to be unimportant \cite{alfa4}. (The one photon exchange approximation
means that the interaction between lepton and target particle is via the 
one photon
exchange). The example of such higher order corrections is
the double bremsstrahlung process depicted in fig.~1b. 
For the scattering on the heavy target there are the other type of corrections
to the elastic radiative tail which should be considered. The effective 
coupling between virtual photon and heavy target particle is very strong due to the
high electric charge $Ze$ of the heavy nucleus. Therefore the one photon
exchange approximation may no longer be valid and the multiphotons exchange
processes should be taken into account as well as the single photon ones.
These multiphotons processes give the contribution which is formally the 
higher order corrections (e.g. $Z^2\alpha^4$ for the double photon exchange
processes) but due to the extra factor $Z$ this contribution can be 
signifficant in comparison with the dominant
Born contribution ($Z \alpha^3$). Naively speaking for the $Z$ large enough
these contributions can be the same order of magnitude. For example: for the
scattering on lead where
$Z\alpha \approx 0.5$, $Z^2\alpha^4$ (double photon exchange contribution)
 is equal to $0.5\cdot Z\alpha^3$. 

\medskip
For the reasons which will be explained in more details in the
next section the estimation of these contribution requires the 
approach different from the usual used Feynman diagrams technique based on the 
expansion in the number of the emitted and exchanged photons (the expansion 
in the power
of $\alpha$).

\medskip
The paper is organized as follows. In section 2 the main idea 
of the approach used is presented. In section 3 the new formula for the 
radiative elastic tail cross section is given and the recalculation
formula for the new 
radiative correction factors $\eta$ is discussed.
Sections 4 and 5 contain the numerical results 
and a 
short summary. The form factors for 
tin and lead and other useful formulae are given in appendices A and B.

\medskip\medskip\medskip
\section{Theory of bremsstrahlung without use of the one photon exchange 
(Born) approximation}

\medskip
As it was mentioned in the previous section the estimation of the elastic
tail in the case of the heavy nucleus target needs a special treatment due
to the fact that the effective coupling is quite strong ($Z\alpha $).
It force us to take into account not only the one photon exchange processes
(presented on fig.~1a) but also more complicated interaction between the
lepton current and the nucleus. In terms of Feynman diagrams such processes are described as
two-, three- and more
photons exchange processes. The example of such processes in the case of two
photons exchange interaction is presented in fig.~1c.

\medskip
The first difficulty which occurs in such approach is the problem of the
description of the intermediate state of the heavy nucleus. In terms of
the Feynman rules it is the question how to describe "nucleus propagator".
The problem is really complicated and difficult to solve in general
(see for example \cite{disper}). Fortunately, in the case of the very
heavy nuclei we can neglect the fact that
the nucleus is free to move when it is struck by virtual photon. Moreover,
we can assume that the interaction between the lepton and the nucleus is 
described by the
electrostatic potential generated by the static nuclear charge distribution.
This is a very good approximation which allows us to avoid the problem
of the nucleus propagator. It corresponds to assume that the mass $M$ 
of the heavy nucleus is infinite.
The corrections to such approximation are
of the order of 1/$M$ and in fact are negligible.
The static potential approximation simplifies the problem and instead of
the diagrams in fig.~1a and fig.~1c we can consider the set of diagrams 
presented in fig.~2. The crosses mean that the interaction is a potential
type and each cross corresponds to the electrostatic form factor function 
$F(q)$ in the relevant matrix element. The form factors take into account 
the nuclear electrostatic structure and are calculated from measured charge
densities \cite{Sick}. The two diagrams in fig. 2a represent the 
famous Bethe-Heitler formula for bremsstrahlung process modified by 
nuclear form factor.

\medskip
The set of diagrams in fig.~2 can 
a priori be calculated to obtain the formula for the elastic tail in the next,
double photon exchange approximation. The calculations are however a little
tricky. Because of the fact that the double photon exchange diagrams (fig. 2b)
lead to an infrared divergent expression for the Coulomb-like potential (1/$r$),
the Yukawa-type ($e^{-\lambda r}$/$r$) should be consider instead. This
allows us to calculate the contribution from subprocess in fig. 2b 
and this contribution is
finite now.
The regularization parameter $\lambda$ appears only in the phase factor of the
matrix element; is not present in the final expression for the cross
section and can be taken to be equal to zero
(see for example \cite{Jauch}). 
The similar problem of the electromagnetic corrections to the process
of the radiative scattering of pions in the Coulomb field of nuclei
has been considered in \cite{pion} and the double photon exchange subprocesses
were taken into account as well as the other higher order electromagnetic
corrections. 

\medskip
The electrostatic field from heavy target nuclei used in the NMC experiment 
is however too strong (tin and lead,
$Z\alpha  \approx 0.5$) to expect that the double photon exchange 
approximation  
will be adequate \footnote{In \cite{pion} the pion scattering on carbon
was considered; here $Z\alpha \approx 0.09$ which justifies the double photon
exchange approximation.}. Moreover, in the case of the very strong field
the perturbation theory based on the expansion in the number of exchanged
photons between nuclei and lepton current is not longer valid and the
different approach is needed.

\medskip
The problem of the scattering in the strong electrostatic field is a very 
old one and was extensively discussed in many papers in the fifties and sixties.
\cite{strong}. The excellent solutions have been found which now are
presented in the textbooks as "classical" exercices. 

\medskip
The radiative processes (bremsstrahlung) in the strong 
fields 
as well as pair production 
were also studied.
A recent application of the old
solutions
of these problems in a new context is the radiation
in the future high energy linear $e^{+}e^{-}$ colliders (so called 
"beamstrahlung" \cite{beam}).

\medskip
The approach to the elastic tail problem presented in our
paper is based on the method discussed in an excellent series of papers by
H.A. Bethe, L.C. Maximon, A. Nordsieck and H.Davies \cite{Bethe}.
The basic ideas are as follows. The real photons emissions from lepton
current line (bremsstrahlung) are treated perturbatively which means that 
only a one real photon emission is cosidered (the lowest
order in $\alpha$). The possible real multiphotons emissions 
the higher order
corrections in $\alpha$ and for the case of the elastic tail
to the deep inelastic scattering are unimportant.
The interaction between leptons and nuclei is described by the Coulomb-like
potential and is {\it not decomposed} into the series of diagrams with 
one-, two-, ... exchanged photons. Instead the new wave function of
the lepton (non-free as in the usuall perturbation theory) is used.
This non-free lepton wave function is calculated as the solution of the 
Dirac equation in the external Coulomb field. 
In this approach the radiative elastic tail is described in the Feynman
diagrams language only by one, simple graph, fig.~3a.
The crossed solid line denotes the modified (non-free) lepton current.
If it would have been possible to solve the Dirac equation exactly, the crossed
solid line in fig.~3a (new lepton wave function) would have been equivalent 
to the
series of the diagrams in the standard perturbative method as it is shown
in fig. 3b. Unfortunately it is not possible to solve the Dirac equation
even if the external potential is a pure Coulomb like. The reason is
that in the relativistic
theory the separation is possible only in the polar coordinates while the
solution in parabolic coordinates is needed in the scattering problem.
The separation of variables in the parabolic coordinates is possible only 
in the 
nonrelativistic case and therefore the approximate solution for the wave
function has to be used. A more detailed discussion can be found
in the original papers \cite{Bethe}.

\medskip
To find the approximate solution of the Dirac equation for the new lepton
wave function in the Coulomb field generated by the heavy nucleus, 
the high energy approximation has been used and the so called "modified 
Furry wave
function" has been obtained. Then, this non-free wave function has been 
used to calculate a matrix element for the bremsstrahlung process from fig.~3a
and finally the cross section formula has been found.
The lost terms in the cross section (due to high energy approximation)
are of the order of $m$/$E^{\prime}$ where $m$ and $E^{\prime}$ are scattered 
lepton mass 
and energy, respectively.

\medskip 
For the high energy deep inelastic muon experiment (NMC) this approximation
works very well and the formula obtained for bremsstrahlung cross section
can be used to estimate the elastic radiative tail
for the case of heavy target scattering. 
However, the solution of the Dirac equation
(Furry wave function) can be calculated analytically only in the case of the
pure Coulomb potential. The real nuclear charge density
modifies the field and in consequence the wave function.
Fortunately, due to the factorization property, it is possible to
take into account the nuclear charge density in the final cross section formula 
via a form factor, but it is done in the inconsistent way.
The factorization and the validity of this approximation will be 
discussed in the next sections.

\medskip\medskip\medskip
\section{The radiative elastic tail cross section.}

\medskip
In this section the main results of Bethe et al. \cite{Bethe} and the
basic formulae for the elastic tail cross section are presented.

\medskip
We start from the well known Bethe-Heitler
formula for the bremsstrahlung in the Born approximation (fig. 2a). Keeping in
mind the future application this formula is given in a little nonstandard way:
\begin{equation}
\frac{d^2\sigma _{BH}}{d\nu d\Omega}=-\frac{Z^2 \alpha^3 |\vec{p^{\prime}}|}
{4\pi^2 |\vec{p}|t }
\int_{q^2_{min}}^{q^2_{max}}\int_{z_{1}}^{z_{2}}\frac{F^2(q)}{q^2} f(z,q^2)dzdq^2 \\
\label{BH}
\end{equation}
with
\begin{equation}
f(z,q^2)=\frac{1}{\sqrt{(z-z_{1})(z_{2}-z)}}
\left[ \frac{4}{q^2} + \frac{S(q^2)}{2 z z^{\prime}} + \frac{m^2}{z^2}\biggl(\frac{4 E^{\prime}^2}{q^2} -1\biggr)
 + \frac{m^2}{z^{\prime}^2} \biggl(\frac{4 E^2}{q^2} -1\biggr)
\right] 
\end{equation}
where: 
\begin{displaymath}
\begin{array}{lll}
\nu = E-E^{\prime}, & \;\;\;\;&
\Omega = (\Theta,\varphi), \\  
p = (E,\vec{p}), & \;\;\;\;&
p^{\prime} = (E^{\prime},\vec{p^{\prime}}),  \\
k = (\omega,\vec{k}), & \;\;& 
Q^2 = - (p-p^{\prime})^2, \\
z = p\cdot k, & \;\;\;\;&
z^{\prime} = p^{\prime}\cdot k,  \\ 
t = \sqrt{\omega^2 + Q^2}, &&  
\end{array}
\end{displaymath} 
and the function $S(q^2)$ is defined as:
\begin{equation}
S(q^2) =  q^2 + \frac{Q^4-16 m^2 E E^{\prime}}{q^2} + 4 m^2 - 4 (E^2 +
E^{\prime}^2) \\
\end{equation}
There is also a relation between $z$ and $z^{\prime}$:
\begin{equation}
z^{\prime} = z + \frac{1}{2}(Q^2-q^2) \nonumber \\
\end{equation}
$p$, $p^{\prime}$ are the momenta of an incoming and outgoing lepton,
$\Omega$ the scattering angle;
$k$ denotes a four-momentum of the emitted real photon. 
Observe, that
$\omega$ is equal to $\nu$ in the potential
approximation. 
The function $F(q)$ is the form factor of the heavy nucleus.
In the case of a pure Coulomb potential $F(q)$ is equal to 1.
More detailed discussion and the relevant formulae for the tin and lead 
form factors
used in NMC analysis is presented in appendix A.

\medskip
The limits of integration, $z_{1}$ and $z_{2}$, are defined by the following
relations:
\begin{eqnarray}
z_{1}+z_{2} &=& \frac{H(q^2)}{2 t^2} \nonumber \\
z_{1} z_{2} &=& \frac{N(q^2)^2}{4 t^2} \nonumber \\
\label{z12}
\end{eqnarray} 
where:
\begin{eqnarray} 
H(q^2) &=& (Q^2+q^2)(E^2-E^{\prime}^2) - (Q^2-q^2)(Q^2+{\omega}^2) \nonumber \\
N(q^2) &=& \sqrt{|\vec{p}|^2 (q^2-q_{p}^2)^2+4 m^2 \omega^2 q^2} \nonumber \\
q_{p}^2 &=& \frac{|\vec{p^{\prime}}| Q^2}{|\vec{p}|} 
\label{NH}
\end{eqnarray}

\medskip
The $q^2$ integration variable has a meaning of the square of the three-momentum
transfer from the lepton to the nuclei \footnote{It is well known that
the energy is not transfered to the static, classical potential field and 
therefore q is a three- instead of a four-momentum vector} and should be correctly written as
 $|\vec{q}|^2$. To simplify the notation we use
$q^2$ instead. 
The integration limits for $q^2$ are the following:
\begin{equation}
q_{min}^2 = (t-\omega)^2,\;\;\;\;\;\;\;\;\;
q_{max}^2 = (t+\omega)^2
\nonumber 
\end{equation}

\medskip
The integration over $z$ in the formula (\ref{BH}) is simple and the final
Born (one photon exchange) formula for the differential cross section
for the elastic tail in the potential approximation can be written as:

\begin{equation}
\frac{d^2 \sigma }{d\nu d\Omega} = -\frac{Z^2 \alpha^3 |\vec{p^{\prime}}|
}{2 \pi |\vec{p}|} \int_{q_{min}^2}^{q_{max}^2}
\frac{F^2(q)}{q^2} \left \{ \frac{2}{t q^2} + \left [\frac{S(q^2)}{(Q^2 - q^2)N(q^2)} +
 \frac{m^2 H(q^2)}{ N(q^2)^3}\right] - 
\biggl[
E \leftrightarrow E^{\prime} 
\biggr] \right \}dq^2
\label{BHC}
\end{equation}

\medskip
The "exact" formula (i.e. without potential but in Born approximation)
fig.~1a,
presented in \cite{MT} and in the second reference in \cite{Dubna}, differs
from (\ref{BHC}) by extra terms proportional to the invers of the mass $M$
of the target particle. The reader can easily check that adding 
the following terms
to the integrand of (\ref{BHC}):
\begin{equation}
\frac{-2F^2(q)}{M N(q^2)} \left \{ \left [ \frac{E}{q^2 } + \biggl(1-\frac{E^{\prime}^2}{q^2}\biggr) \frac{2 m^2 Q^2 (E+E^{\prime})}{N(q^2)^2} 
+ \biggl(1-\frac{2 m^2}{q^2}\biggr) \frac{\omega(|\vec{p}|+|\vec{p^{\prime}}|)}
{(Q^2-q^2) (E+E^{\prime})}\right ]  +
\biggl [ 
E \leftrightarrow E^{\prime} 
\biggr] \right \} 
\end{equation}
and
\begin{equation}
\frac{2 m^2F^2(q)}{M^2 N(q^2)} \left \{ \left [\frac{(|\vec{p}|+|\vec{p^{\prime}}|)}
{(Q^2-q^2)
(E+E^{\prime})} + Q^2 E^{\prime}\frac{(E+E^{\prime})}{N(q^2)^2} \right ]
-\biggl [ 
E  \leftrightarrow E^{\prime} 
\biggr] \right \} 
\end{equation} 
restores the elastic tail cross section formula from \cite{MT,Dubna}
\footnote{$q^2$ should be replaced by $- q^2+\frac{q^4}{4 M^2}$ where
now $q^2$ is a real four-momentum vector transmitted to the target.}.
These terms are however negligible for heavy targets.

\medskip
Let us now reffer to the main result obtained by Bethe et al. in \cite{Bethe}.
This result is the key point in our further considerations.
As it was said in the previous section the non-free modified Furry
wave function has been used to determined the cross section for the 
bremsstrahlung
without Born approximation (fig.~3a). It was found to be the Bethe-Heitler
one photon result {\it multiplied by the factor $R$ } which depends on the 
$z$ and 
$q^2$. This fact can be symbolically depicted as in fig.~3c.
This result was obtained under the assumption that the potential field
generated by the nucleus is a pure Coulomb potential ($F(q) = 1$). 
Unfortunately the Dirac equation for
the wave function in more realistic shape of the potential
cannot be solved analytically.
Therefore we propose to put the nucler charge density in the final
formula for the cross section in the following way: use factor $R$ calculated
for the pure Coulomb potential and improve Bethe-Heitler one photon part of the
formula
by  multiplying it by the square of the nuclear form factor $F^2(q)$, as 
in the case of the Born approximation. 
Let us note that the form factor
for the heavy nuclei is strongly peaked up near $q^2\approx 0$.
Therefore the main contribution to the cross section follows
from the configuration with smallest $q^2$ when the field is very similar to
the pure Coulomb field. It means that the using of the factor $R$ calculated
for the pure Coulomb potential is reasonable.

\medskip
Our new "master" formula for the elastic radiative tail (bremsstrahlung) for
the heavy targets can be expressed as follows:
\begin{equation}
\frac{d^2\sigma _{tail}}{d\nu d\Omega}=-\frac{Z^2 \alpha^3 |\vec{p^{\prime}}|
}{4\pi^2 |\vec{p}|t }
\int_{q^2_{min}}^{q^2_{max}}\int_{z_{1}}^{z_{2}}\frac{F^2(q)}{q^2} R(z,q^2) 
f(z,q^2)dzdq^2 \\
\label{NBH}
\end{equation}
The factor function $R$ obtained in \cite{Bethe} is:
\begin{equation}
R(z,q^2) = \frac{V^2(x)+(Z\alpha)^2 (1-x)^2 W^2(x)}{V(1)^2}
\label{R}
\end{equation}
where:
\begin{eqnarray}
V(x) &=& {\cal{F}}(-i Z\alpha,iZ\alpha,1,x) \nonumber \\
W(x) &=& \frac{1}{(Z\alpha)^2}\frac{dV(x)}{dx} \nonumber \\
x &=& 1-\beta z z^{\prime} \nonumber \\
\beta &=& 1\over{q^2 E E^{\prime}}
\label{zx} 
\end{eqnarray}
and
\begin{equation}
V(1) = \frac{\sinh (\pi Z\alpha)}{\pi Z\alpha}
\end{equation}
The function $\cal{F}$ is a hypergeometric function.
The $x$ is tending to $0$ when $q^2$ is very close to minimal value $q^2_{min}$
and never exceeds $1$.
It is important to notice that the function $R$ is also
less than 1 and monotonically increases with $x$.
It means that the bremsstrahlung cross section (\ref{NBH}) is 
{\it always}
smaller than the Bethe-Heitler cross section in the Born approximation 
(\ref{BH}).
Due to 
the fact that the
nuclear form factor enhances small $q^2$ configurations ($x$ near $0$)
the difference
between the Born cross section can be significant~\footnote{In contrast to atom
form factor which kills small $q^2$ configurations (complete screening limit)
and in that case the Born Bethe-Heitler formula describes {\it whole} 
bremsstrahlung contribution.}.
From these considerations it is seen that the nuclear form factor $F(q)$ 
plays an important role and must be taken into account.
The integration over $z$ and $q^2$ in (\ref{NBH}) is complicated and will
be discussed in the next section in detail. 

\medskip
We shall close this section with the formula which allows to recalculate
radiative correction factor $\eta$ including presented approach
to the elastic tail. 
The $\eta$ factor will be changed only in that kinematical region
where the elastic radiative tail correction is important; for
the small $x_{Bj}$ and high $y$. The detailed description 
of the  
the different types of the corrections to the deep inelastic muon 
scattering
can be found for example in \cite{my}.
Let $\eta_{old}$ denote the radiative correction factor for the radiative 
correction scheme where the elastic tail is treated in the Born
approximation. Then the new factor, $\eta_{new}$ is:
\begin{equation} 
\eta_{new} = \frac{\eta_{old}}{1+\eta_{old} \frac{\Delta \sigma_{tails}}{
\sigma_{1\gamma}}}
\label{eta}
\end{equation}
where:
\begin{equation}
\Delta \sigma_{tails} = \sigma_{new}^{tail}-\sigma_{old}^{tail} \nonumber \\
\end{equation}
The subscript "new" reffers to the cross section for the elastic
tail calculated by the formula (\ref{NBH}). The $\sigma_{1\gamma}$ is the
one photon cross section for the deep inelastic electroproduction (see e.g. \cite{my}).

\medskip\medskip\medskip
\section{Numerical calculations}

\medskip
The new formula for the heavy target radiative elastic tail cross section
(\ref{NBH}) must be integrated over $z$ and $q^2$ variables. In the contrast to
the Bethe-Heitler cross section (\ref{BH}) it is not simple to integrate over
$z$ because of the extra factor $R$. In the original paper \cite{Bethe} the 
method of integration has been discussed but it requires using of a set of
the angular variables instead of $z$ and $q^2$. 
As the form factor $F(q)$ is rather complicated function (see appendix A)
in our case one of the
variables should be choosen as $q^2$ which
practically excludes
the method proposed in \cite{Bethe}. 

\medskip
The one possibility is to integrate the formula (\ref{NBH}) numerically over
$z$ and $q^2$. The numerical integration over $z$ however can be dangerous and
difficult because the integrand is a product of $R$ factor (which contains 
nontrivial
combination of the hypergeometric functions) and the function $f$ which is
divergent in the limits of the integration. Moreover, such double numerical
integration \footnote{three dimensional integration in the case of tin,
where the form factor must be also calculated numerically, see appendix A.}
is the computer time consuming if the calculations
have to be repeated many times as it is done in the radiative 
corrections analysis. Therefore the semianalytical aproach is proposed
which allows us to avoid the numerical difficulties with integration over $z$.

\medskip
The $R$ factor contains the hypergeometric functions $V(x)$ and $W(x)$
(\ref{R}). 
These functions are convergent inside
the circle $|x|<1$ which allows to approximate the $R$ factor   
by the polynomial in the $x$ variable.
However there is no method to determin {\it a priori} which order of 
the polynomial is sufficient for getting
the reasonable approximation. It must be checked at the end by
comparing the results obtained with the different order polynomials used
instead of the exact form for $R$ factor.
It is a nontrivial problem because the $W(x)$ function is converging very
slowly, especially for $x$ near 1.
Keeping in mind that $x$ is a simple function of $z$ variable (\ref{zx}) we 
have for $R$:  
\begin{equation}
R = \frac{1}{V(1)^2} \sum_{i=0}^{n} C_{i} \left [ 1-\frac{1}{2}(Q^2-q^2) \beta z -
\beta z^2\right ]^{i} = \frac{1}{V(1)^2} \sum_{i=0}^{2n} g_{i}(q^2) z^{i} \\
\label{ER}
\end{equation}
and the formula for the integrand from \ref{NBH} can be rewritten as follows:
\begin{equation}
-\frac{Z^2 {\alpha}^3 |\vec{p^{\prime}}|}{4{\pi}^2 V(1)^2 t |\vec{p}|} \frac{
F^2(q)} 
{q^2} f(z,q^2) \sum_{i=0}^{2n} g_{i}(q^2) z^{i} \\
\end{equation}
The numerical coefficients $C_{i}$ depend on the target; are the functions of 
$Z\alpha$.
Now, the integration over $z$ can be performed analytically and 
the final formula
for the elastic tail cross section is in the form of the single integral over
$q^2$ possible to integrate numerically without any difficulties:
\begin{eqnarray}
\frac{d^2\sigma _{tail}}{d\nu d\Omega}&=&-\frac{Z^2 \alpha^3 |\vec{p^{\prime}}|
}{4\pi^2 |\vec{p}|t V(1)^2}
\int_{q^2_{min}}^{q^2_{max}} \frac{F^2(q)}{q^2} \sum_{i=0}^{2 n} \left [
\frac{1}{2}S(q^2) I_{-1}^{i-1} + 
m^2 \biggl(\frac{4 E^{\prime}^2}
{q^2} -1\biggr) I^{i-2} \right. \nonumber \\ &&\;\;\;\;\;\;\;\;\;\;\;\;\;
\;\;\;\;\;\;\;\;\;\;\;\;\; \left. + \frac{4}{q^2} I^{i}
+ m^2 \biggl(\frac{4 E^2}{q^2} -1\biggr) I_{-2}^{i}\right ]
g_{i}(q^2) dq^2 
\label{FT}
\end{eqnarray}
The functions $I_{-2}^{i}$, $I_{-1}^{i}$ and $I^{i}$ depend on 
$q^2$
The full expressions
for these functions are given in the appendix B.

\medskip
To calculate the expansion functions $g_{i}(q^2)$ the {\it Mathematica}
program for symbolic computations was used. As it was tested
the satisfactory $n$ order used in the expansion should be greater that 20,
so we put $n = 23$ in the calculations.
The {(\it analytically)} obtained functions were rewritten and used in 
the Fortran code. The formula (\ref{FT}) is very long but the integration
is very fast.

\medskip
The numerical integrations were performed for two heaviest elements
used as targets in the NMC experiment: tin and lead. 
We also repeated the calculations for calcium and found that the
results are practically exactly the same as in the Born 
approximation. For tin and lead the small differences
were found which affect the total $\eta$ correction factor for the region
of the smallest values of $x_{Bj}$ measured in the experiment.
The new $\eta$ correction factor was calculated for four different muon
energies: 90, 120, 200 and 280 GeV for the tin target, and for 200 and 280 GeV 
for lead.
(all energies at which the data were collected for these targets 
in NMC experiment).  

\medskip
In figs 4 and 5 the differences between "new" and "old" $\eta$ factors
are presented for different values of $x_{Bj}$ and $y$ for tin and lead,
respectively. As the reference,
$\eta_{old}$, the $\eta$ correction factor calculated in the frame of the 
Dubna scheme by the code TERAD86 \cite{Dubna}, used in NMC data 
analysis, has been 
taken. The "new" tail is from 1.5 \% (tin) to 3~\% (lead) smaller than
calculated in one photon approximation and the difference between cross
sections for the elastic tails is practically independent off
the $x_{Bj}$ and $y$. The strong dependence on $x_{Bj}$ and $y$ 
of the difference
in $\eta$ factor which is seen in figs 4 and 5, is a reflection of the 
importance of the elastic tail correction in total correction factor.
As it was mentioned above, the elastic tail contribution is important
in the small $x_{Bj}$ region and mainly for high~$y$.

\medskip\medskip\medskip
\section{Summary}

\medskip
In this paper the method of calculations of the
bremsstrahlung process in the strong electrostatic field without
the use of the Born approximation is presented. The method originally
proposed by Bethe et al. \cite{Bethe} was modified and applied to improve the
radiative correction scheme used in the deep inelastic muon scattering
data analysis for heavy targets.

\medskip
The results show that the 
Born approximation which is a standard approach in the
elastic tail calculations, used in
all radiative correction schemes \cite{MT,Dubna}, is not longer valid 
for the case of the
scattering on very heavy targets like tin and lead. In that cases the
small corrections (up to 3\%) were found in the small $x_{Bj}$ region.
The effects of such size might be
important for the very precise 
deep inelastic high energy muon scattering experiment NMC \cite{NMC}.

\medskip
The method described in the paper is also based on certain approximations
which
can be the sources of the uncertainties in the calculations. However we 
think that
the used approximations are well justified.
We would like
to close this paper with some remarks about the uncertainties.
There are three sources of them: the first connected with the high energy
approximation, the second with the fact that we put the form factors
in the final formula in inconsistent way, ignoring it during the calculations of
the Furry wave function and - in consequence in the factor $R$ and third,
following from the terms lost in the expansion of $R$ in numerical
calculations. The high energy approximation is detaily discussed in the
original paper \cite{Bethe} and the error is estimated to be of order
of $m$/$E^{\prime}$ and can be significant for the smallest muon energy
and the highest $y$.
The lost terms in $R$ can increase the elastic radiative tail cross section
and reduce the difference between Born cross section and our calculations.   
To check it $n= 50$ was tried and the results differs less than 0.5 \%. 
It is very difficult to estimate the error from $R$
factor calculations when the pure Coulomb potential was assumed.
Taking into account that the form factor 
enhances the smallest $q^2$ configurations which
are mainly responsible for the fact that presented result
differs from Born approximation, the lack of the form factor in the part
of calculations suggests that the cross section
can be a little overestimated.
This error seems to be work in the opposite direction than 
previous one. 
In another words it is a question of the usefulness of the Furry's
wave function. It is certainly correct as a first approximation.

\medskip \medskip \medskip
\noindent
{\Large {\bf {Acknowledgements}}} \\

\medskip
I would like to thank Antje Br$\ddot{u}$ll for help in the calculations.
I also thank Barbara Badelek, Ewa Rondio and 
my colleagues from the NMC.
\newline
This research was supported in part by the Polish Committee for Scientific
Research, grant number  2 P302 069 04.

\medskip\medskip\medskip
\section{Appendix A}

\medskip
The detailed description of the methods of extracting nuclear charge densities
from the experiments the reader can find in \cite{Sick}. The relation between
density $\rho (r)$ and the form factor $F(q)$ is a Fourier transform:
\begin{equation}
F(q) = \int_{-\infty}^{\infty} \frac{\rho (r)}{Z e} e^{-\vec{r}\cdot \vec{q}}
d^3 r  \\
\end{equation}
where the normalization condition used for $\rho$ is:
\begin{equation}
\int_{0}^{\infty}4\pi r^2 \rho (r) dr = Ze  \\
\label{nor}
\end{equation} 

\medskip
The so called SOG (Sum of Gaussians) method has been used in determination
 of the
lead and calcium charge density. In that method the charge density $\rho$ is 
assumed to be
in the form:
\begin{equation}
\rho (r) = \frac{Ze}{2{\pi}^{3/2} \gamma^3} \sum_{i=1}^{N} \frac{Q_{i}}{1+
\frac{2 R_{i}^2}{\gamma^2}} (e^{-(r-R_{i})^2/\gamma^2} + e^{-(r+R_{i})^2/\gamma^2})
\end{equation}
The SOG method allows to calculate the form factor analytically:
\begin{equation}
F(q) = e^{-\frac{1}{4} q^2 \gamma^2} \sum_{i=1}^{N} \left[ \cos(qR_{i}) +
\frac{2 R_{i}^2}{\gamma^2} \frac{\sin(qR_{i})}{qR_{i}} \right] \\
\end{equation}
The $Q_{i}$, $R_{i}$ and $\gamma$ parameters are determined from {\it elastic}
scattering experiments on lead. In the radiative correction scheme
used in NMC analysis the 12-parameter SOG method has been used to determined
the form factor for lead and the values of parameters have been taken from
\cite{lead}.

\medskip
The 3-parameter Gauss approximation has been used for tin charge density:
\begin{equation}
\rho(r) = \frac{Ze}{4\pi C_{0}} \cdot \frac{1+\frac{w}{c^2}r^2}{1+e^{(r^2-c^2)
/z^2}} 
\end{equation}
To determine the tin's form factor  
the Fourie transform of the tin's charge density had to be numerically
calculated. 

\medskip
The NMC tin target was a mixture of 9 tin isotopes: 112 (1\%), 114 (0.7\%),
116 (14.7\%), 117 (7.7\%), 118 (24.3\%), 119 (8.6\%), 120 (32.4\%), 122 (4.5\%)
and 124 (5.6\%). The small admixture of Sn 115 isotope was neglected in 
calculations.
To calculate the form factor the averaged charge density has been taken.
The values of $c$, $w$ and $z$ parameters for each isotope have been determined 
from tin elastic 
scattering data and taken from \cite{cyna}. 
The normalization constant $C_{0}$ can be calculated via normalization
condition (\ref{nor}).

\medskip\medskip\medskip
\section {Appendix B}

\medskip
Below we list the integrals $I$ used in the final formula (\ref{FT}):
\begin{equation}
I_{j}^{n} = \int_{z_{1}}^{z_{2}} \frac{z^n z^{\prime}^j}{\sqrt{(z-z_{1}) (
z_{2}-z)}} dz \\
\end{equation}
and
\begin{eqnarray}
I^{n} &\equiv& I_{0}^{n} \nonumber \\
I_{n} &\equiv& I_{n}^{0} \nonumber \\
I^{0} &=& \pi \nonumber \\
I^{1} &=& \pi \frac{(z_{1}+z_{2})}{2} \nonumber \\
I^{-1} &=& \pi \frac{2 t }{N(q^2)} \nonumber \\
I^{-2} &=& \pi \frac{2 t H(q^2)}{N^3(q^2)} \nonumber \\
I_{-1} &=& \pi \frac{2 t }{L(q^2)} \nonumber \\
I_{-2} &=& \pi \frac{2 t}{L^3(q^2)} \left [H(q^2) + 2 (Q^2-q^2) t^2 \right ]\\ 
I_{-1}^{-1} &=& \frac{2}{Q^2-q^2} (I^{-1} - I_{-1}) 
\end{eqnarray}
The integrals $I_{j}^{n}$ for higher indecies $n$ can be calculated 
recurrently:
\begin{eqnarray}
I^{n} &=& (1-\frac{1}{2n}) (z_{1}+z_{2}) I^{n-1} - (1-\frac{1}{n})(z_{1}z_{2}) I^{n-2} \nonumber \\
I_{-1}^{n} &=& I^{n-1} -\frac{(Q^2-q^2)}{2} I_{-1}^{n-1} \nonumber \\
I_{-2}^{n} &=& I_{-1}^{n-1} - \frac{(Q^2-q^2)}{2} I_{-2}^{n-1} 
\end{eqnarray}
Finally, the $L(q^2)$ is equal to $N(q^2)$ 
if $E \leftrightarrow E^{\prime}$. 

\medskip \medskip \medskip
\noindent
{\Large {\bf {Figure Captions}}}
\medskip \medskip
\begin{enumerate}
\item Feynman diagrams for the radiative elastic tail 
in the one photon
exchange approximation: a) - Born approximation,
b) - example of 
the higher order corrections. 
The double photon exchange subrocesses 
contributing to the radiative elastic tail are shown in c).

\item Feynman diagrams for the radiative elastic tail 
in the potential approximation:
a) - Born approximation, 
b) - double photon exchange
approximation.

\item Feynman diagrams for the radiative elastic tail for the heavy
target without use to one photon approximation ( a)) 
The crossed solid line refers to
non-free modified Furry's lepton wave function, which formally can
be decomposed into series of standard perturbation theory diagrams ( b)).
The result of Bethe et al. \cite{Bethe} is symbolically presented in c).

\item The ratio of the total correction factors $\eta$ for tin target, 
calculated for different $x_{Bj}$ and $y$ and for 
four different energies: 90, 120, 200 and 280 GeV, used in NMC experiment. 
The "new" subscript
denotes that the elastic tail in $\eta$ factor was calculated with the 
method presented in this paper; the "old" one is with one photon Born
approximation. All remaining radiative corrections to the deep inelastic
muon scattering have been calculated using TERAD86 code \cite{Dubna}.

\item The ratio of the total correction factors $\eta$ for lead target,
calculated for different $x_{Bj}$, $y$ and for two energies: 200 and 280 GeV
(as used in NMC experiment).

\end{enumerate}

\newpage
\begin{figure}
\centering
\epsfig{file=fig1.ps}
\end{figure}


\newpage
\begin{figure}
\centering
\epsfig{file=fig2.ps,
bburx=530pt,bbury=730pt}
\end{figure}

\newpage
\begin{figure}
\centering
\epsfig{file=fig3.ps,
bburx=530pt,bbury=730pt}
\end{figure}

\newpage
\begin{figure}
\centering
\epsfig{file=fig4.ps,height=20cm,
bburx=530pt,bbury=730pt}
\end{figure}

\newpage
\begin{figure}
\centering
\epsfig{file=fig5.ps,height=20cm, 
bburx=530pt,bbury=730pt}
\end{figure}

\end{document}